\documentstyle[12pt,aasms4,epsf]{article}
\slugcomment{}

\begin{document}

\newcommand{\Einstein}{{\it Einstein}}
\newcommand{\EXOSAT}{{\it EXOSAT}}
\newcommand{\Ginga}{{\it Ginga}}
\newcommand{\ROSAT}{{\it ROSAT}}
\newcommand{\ASCA}{{\it ASCA}}

\newcommand{\HubbleCNST}{{\it H$_0$}}
\newcommand{\FX}{{\it f$_{\rm X}$}}
\newcommand{\LX}{{\it L$_{\rm X}$}}
\newcommand{\KT}{{\it k$_{\rm B}T$}}

\newcommand{\APJ}{{\it Astrophys.\ J.}}
\newcommand{\APJL}{{\it Astrophys.\ J. Letters}}
\newcommand{\APJS}{{\it Astrophys.\ J. Suppl.}}
\newcommand{\PASJ}{{\it Publ.\ Astron.\ Soc.\ Japan}}
\newcommand{\PASJL}{{\it Publ.\ Astron.\ Soc.\ Japan Letters}}
\newcommand{\MNRAS}{{\it Monthly Notices Roy.\ Astron.\ Soc.}}
\newcommand{\AAP}{{\it Astron.\ Astrophys.}}
\newcommand{\AAPS}{{\it Astron.\ Astrophys.\ Suppl.}}
\newcommand{\ARAA}{{\it Ann.\ Rev.\ Astron.\ Astrophys.}}
\newcommand{\NATURE}{{\it Nature}}

\title{ICM masses in nearby and distant clusters of galaxies}

\author{Takeshi Go Tsuru\altaffilmark{1},
	Hironobu Matsumoto,
	Katsuji Koyama,
	Hiroshi Tomida}
\affil{Department of Physics, Kyoto University, \\
	Kitashirakawa-Oiwake, Sakyo-ku, Kyoto 606-01, JAPAN}

\author{Yashushi Fukazawa}
\affil{Department of Physics, University of Tokyo, \\
	7-3-1 Hougou, Bukyou-ku, Tokyo 113, JAPAN}

\author{Makoto Hattori}
\affil{Astronomical Institute, Tohoku  University, \\
	Aoba, Sendai 980-77, JAPAN}

\and

\author{John P. Hughes}
\affil{Department of Physics, Rutgers University, \\
	136 Frelinghuysen Road, Piscataway, NJ 08854, USA}
	% jph@physics.rutgers.edu 

\altaffiltext{1}{tsuru@cr.scphys.kyoto-u.ac.jp} 

\begin{abstract}
We compile X-ray data of clusters of galaxies 
at a wide range of redshift $z=0.0\sim1.0$ 
and compare X-ray luminosities, gas masses, 
iron abundances and iron masses of nearby clusters and 
those of distant clusters 
as a function of plasma temperature. 
All the date presented in this paper have been obtained 
only with \ASCA\ and/or \ROSAT. 
No evidence for a change in the \KT -\LX\ relationship
or \KT - iron abundance at $z<0.6$ is found.
No significant change in \KT -\LX\ relationship
at $z>0.6$ is seen, either.
However, the gas masses of very distant clusters at $z\sim0.8-1.0$ 
and some clusters around $z\sim0.6$
are significantly lower than those of nearby clusters, 
which indicates a hint of evolution. 
The iron masses of the two very distant clusters at $z\sim0.8-1.0$ are
significantly different from each other, 
which suggests that the epoch of the metal injection into ICM
is different from cluster to cluster.

\end{abstract}


\section{Introduction}
Investigation of evolution of X-ray properties of clusters of galaxies
is a key study of cosmology.
The most important result in this field
before \ROSAT\ and \ASCA\ observatories is
the detection of the negative evolution of the X-ray luminosity function
at red-shift lower than 0.6 (Gioia et al. 1990).
Following it, many groups have been investigating
evolution of X-ray luminosity function from various surveys
with \ROSAT\ observatory (eg. Collins et al. 1997). 
Many of them indicate no negative evolution at the red-shift lower than 0.7,
which is against the \Einstein\ result.

\ASCA\ added new and key information
to the study of the evolution of clusters;
temperature and metal abundance of distant clusters of galaxies.
We review it and show new results in this report.

\section{Results}
\subsection{Temperature and X-ray luminosity}
Tsuru et al. (1996) and Mushotzky and Scharf (1997) compiled
\ASCA\ data of distant clusters of galaxies (mostly $0.6>z>0.1$).
By comparing it with the data of nearby clusters
obtained with the previous observatories (eg. David et al., 1993),
they suggested no evidence for a change
in the temperature and X-ray luminosity relationship.
However, it is still doubtful
because it is shown without enough cross-calibration
among \ASCA\ and the other previous observatories.
Recently, Fukazawa (1997) compiled \ASCA\ data of nearby clusters.
Then, We make comparison with the data and show result in this report,
which is free from the difficulty of the cross-calibration.

\ASCA\ observations of two very distant clusters,
AXJ2019+1127 and MS1054-0321 at the red-shifts of 1.0 and 0.829 respectively,
were reported very recently (Hattori et al. 1997a; Donahue et al. 1997).
Including the two clusters, We show the temperature and
X-ray luminosity relationship in the figure~1.
The figure indicates no significant difference among the epochs,
which implies no evidence for a strong evolution in this relationship.

\placefigure{fig1}

\subsection{Temperature and Gas Mass Relationship}
Next, we compare the two clusters with nearby clusters
in the temperature and gas mass relationship in the figure~2.
We also plot data of the clusters at the red-shift of around 0.5.
All the temperatures plotted in the figure are obtained with \ASCA.
The gas masses for nearby clusters are calculated
from results of imaging analyses of \ASCA\ data (Fukazawa 1997).
The gas masses of the other (distant) clusters are obtained with \ROSAT\
observatory except for that of 3C295 which is determined
with \Einstein\ observation (Hughes and Birkinshaw 1995; Donahue 1996;
Schindler et al. 1997; Henry and Henriksen 1986; Donahue et al. 1997;
Hattori et al. 1997a; Hattori et al. 1997b; Hughes 1997).
All the gas masses except for AXJ2019+1127 are defined
as those within 1.0~Mpc from cluster center.
In the case of AXJ2019+1127, observed $R_{\rm max}$ is 0.5~Mpc.
Then, adding to the gas mass actually detected in $R_{\rm max}$ of 0.5M~pc,
we plot extrapolated gas mass
when assuming $R_{\rm max}$ of 1.0~Mpc in the figure.

\placefigure{fig2}

This figure indicates significant difference
between the nearby and distant clusters.
The gas mass of the two very distant clusters, AXJ2019+1127 and MS1054-0321,
are only 10\% or 25\% of those of nearby clusters
when comparing at the same temperature.
Two other distant clusters at the red-shift of around 0.5,
3C295 and MS0451.6-0305 also contain only
20\%-50\% gas masses of nearby clusters
when comparing at the same temperature.
Thus, we found a hint of evolution in this relationship.

It is suggestive that their gas masses are
much smaller than those of nearby cluster
although the no significant difference is seen
in the temperature and X-ray luminosity relationship.
It should indicate different distribution of ICM between the two groups.
Small amount of gas can emit large luminosity
if its distribution is compact.
The $\beta_{\rm fit}$ of AXJ2019+1127 and MS1054-0321 determined with
\ROSAT/HRI are $\sim 0.9$ and $0.66-1.0$, respectively
(Hattori et al. 1997a; Donahue et al. 1997).
The values are larger than the typical value of
$\beta=0.6-0.7$ for nearby rich clusters,
which indicates the two clusters are compacter than nearby clusters.

\subsection{Temperature and Iron Abundance}
It has been already reported that
no evidence for a change in the iron abundance and temperature relationship
as a function of red-shift is seen at $z<0.6$
(Tsuru et al., 1996;  Mushotzky and Loewenstein 1997).
In this report, we add the new data of MS1054-0321 and AXJ2019+1127
to the relationship in the figure~3.
The iron abundance of MS1054-0321 is consistent with nearby clusters.
However, that of AXJ2019+1127 is extremely higher than the relationship.

\placefigure{fig3}

\subsection{Temperature and Iron Mass Relationship}
Next, we show the temperature and iron mass relationship in the figure~4.
The iron mass of AXJ2019+1127 is consistent
with the relationship of the nearby clusters.
Since its low gas mass and high iron abundance counterbalances each other,
the iron mass comes on the relationship.
In the case of MS1054-0321,
the abundance is consistent with nearby clusters but the gas mass is very low.
Then, the iron mass becomes very low.

\placefigure{fig4}

The result of AXJ2019+1127 indicates
that the metal injection process into its ICM through its life had already
finished before the red-shift of 1.0.
On the other hand, it is suggested that the metal injection in MS1054-0321
has not started yet.

\section{Summary}
(1) No evidence for a change in the \KT -\LX\ relationship
or \KT - iron abundance at $z<0.6$ is found.
(2) No significant change in \KT -\LX\ relationship
at $z>0.6$ is seen, either.
However, the gas masses of very distant clusters at $z\sim0.8-1.0$
and some clusters around $z\sim0.6$
are significantly lower than those of nearby clusters,
which indicates a hint of evolution.
(3) The iron masses of the two clusters at $z\sim0.8-1.0$ are
significantly different,
which suggests that the epoch of the metal injection into ICM
is different from cluster to cluster.

% \placetable{tbl-1}

\acknowledgments

% We are grateful to ASCA team.

% \appendix

\clearpage

\epsfile{file=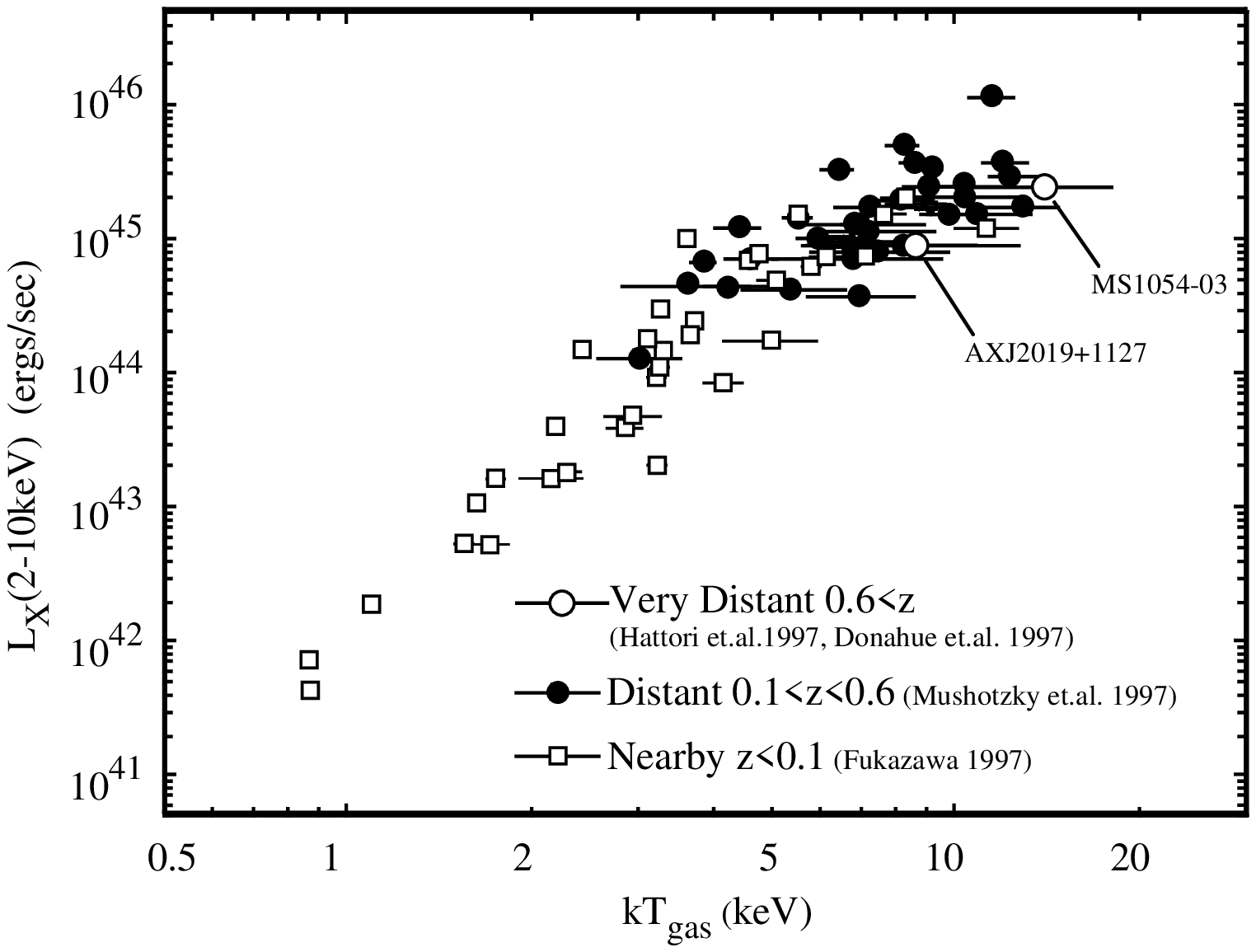,scale=0.70}
\figcaption[kT_LX_z0-1.eps]
	{
	The temperature and X-ray luminosity relationship.
        Since all the data were obtained with \ASCA\ observatory,
        the result is free from difficulty of cross-calibration
        (Fukazawa 1997; Mushotzky and Scharf 1997;
        Donahue et al. 1997; Hattori et al. 1997a). 
	\label{fig1}
	}

\clearpage

\epsfile{file=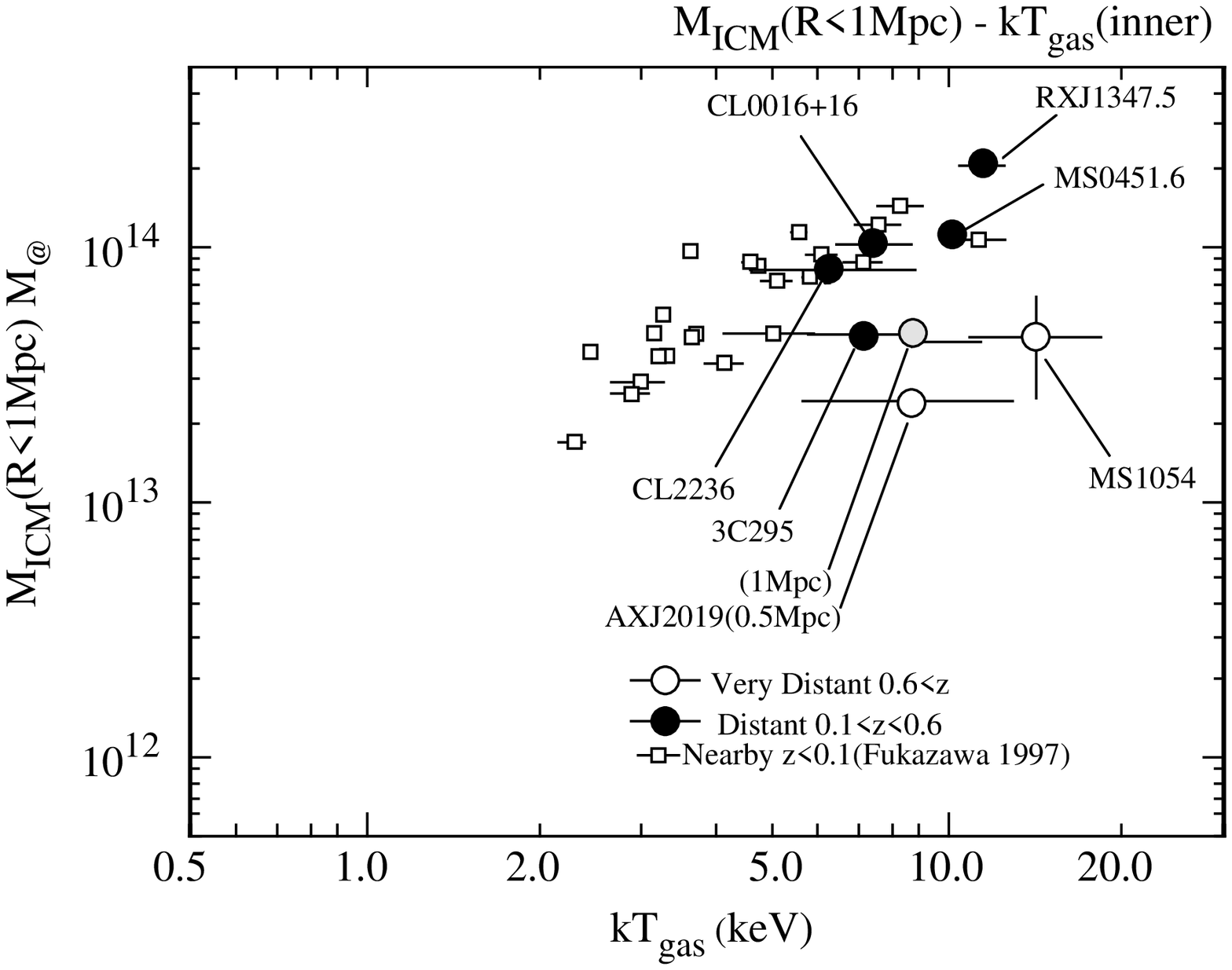,scale=0.70}
\figcaption[kT_Mgas_z0-1.eps]
	{
	The temperature and gas mass relationship. 
	All the temperatures were obtained with \ASCA.
	The gas masses for the nearby clusters were calculated from 
	the results of imaging analysis of \ASCA data (Fukazawa 1997). 
	The gas mass of 3C295 is derived with \Einstein\ result 
	(Henry and Henriksen 1986). 
	The other masses are adopted from \ROSAT\ results 
	(Hughes and Birkinshaw 1995; Donahue 1996; Schindler et al. 1997;
	Donahue et al. 1997; Hattori et al. 1997a; Hattori et al. 1997b;
	Hughes 1997).
	\label{fig2}
	}

\clearpage

\epsfile{file=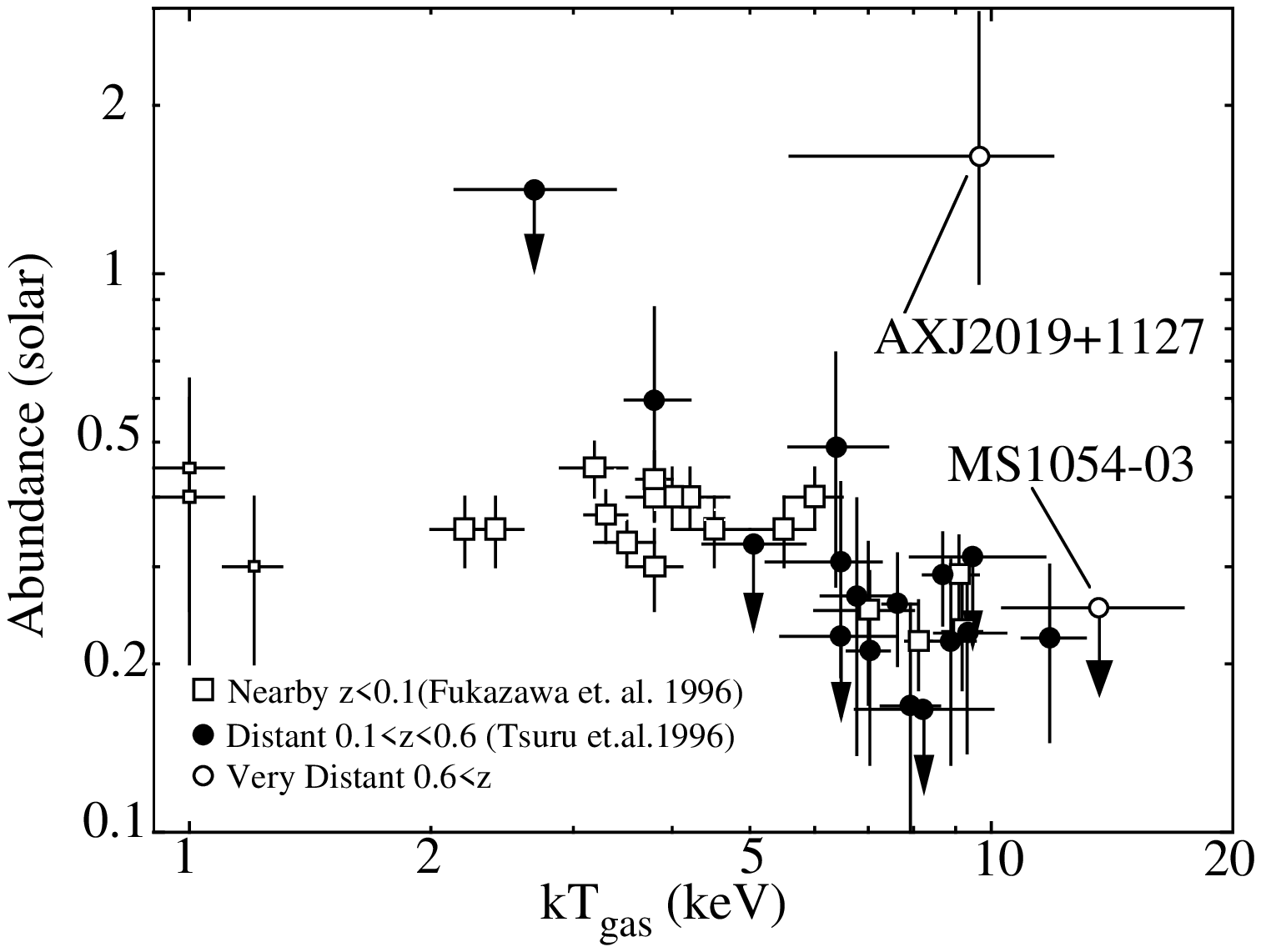,scale=0.70}
\figcaption[kT_Ab_z0-1.eps]
	{
	The temperature and iron abundance relationship. 
	All the data were determined with \ASCA\ 
	(Fukazawa 1997; Tsuru et al 1996; 
	Donahue et al. 1997; Hattori et al. 1997a).
	\label{fig3}
	}

\clearpage

\epsfile{file=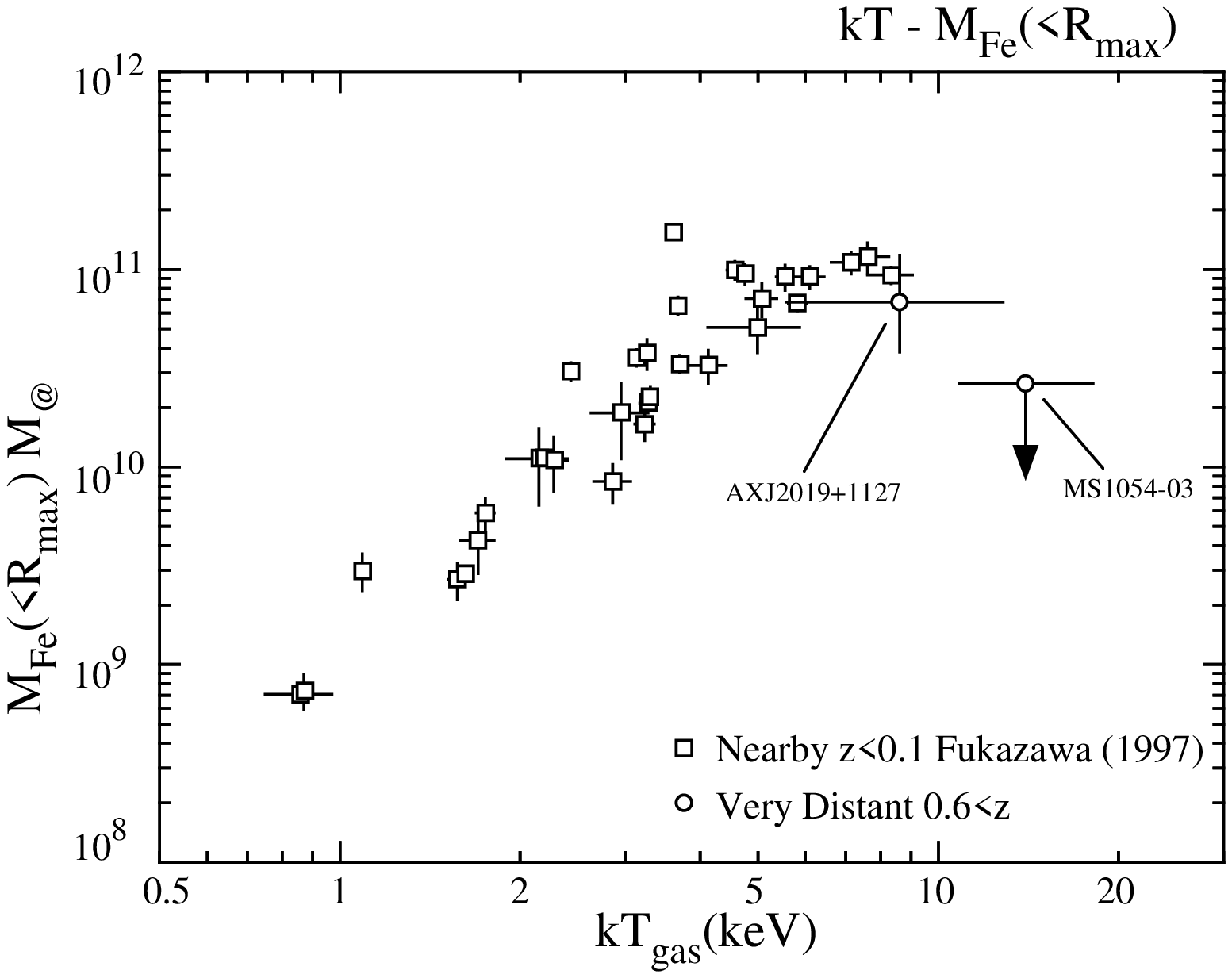,scale=0.70}
\figcaption[kT_MFe.eps]
        {
	The temperature and iron mass relationship.
        The iron mass is defined as that with in $R_{\rm max}$
        (Fukazawa 1997; Donahue et al. 1997; Hattori et al. 1997a).
	\label{fig4}
	}

\end{document}